# Fe-based high temperature superconductivity with $T_c$=31K bordering an insulating antiferromagnet in (Tl,K)$Fe_xSe_2$ Crystals

Minghu Fang[1*], Hangdong Wang[1,2], Chiheng Dong[1],
Zujuan Li[1], Chunmu Feng[1], Jian Chen[1], H.Q. Yuan[1]

[1] Department of Physics, Zhejiang University, Hangzhou 310027, China
[2] Department of Physics, Hangzhou Normal University, Hangzhou 310036， China

**Up to now, there have been two material families, the cuprates[1] and the iron-based compounds[2-12] with high-temperature superconductivity (HTSC). An essential open question is whether the two classes of materials share the same essential physics. In both, superconductivity (SC) emerges when an antiferromagnetical (AFM) ordered phase is suppressed. However, in cuprates, the repulsive interaction among the electrons is so strong that the parent compounds are "Mott insulators." By contrast, all iron-based parents are metallic. One perspective[13-14] is that the iron-based parents are weakly correlated and that the AFM arises from a strong "nesting" of the Fermi surfaces. An alternative[15-18] view is that the electronic correlations in the parents are still sufficiently strong to place the system close to the boundary between itinerancy and electronic localization. A key strategy to differentiate theses views is to explore whether the iron-based system can be tuned into a Mott insulator. Here we identify an insulating AFM in (Tl,K)$Fe_xSe_2$ by introducing Fe-vacancies and creating superconductivity in the Fe-planar. With the increasing Fe-content, the AFM order is reduced. When the magnetism is eliminated, a superconducting phase with $T_c$ as high as 31K (and a $T_c^{onset}$ as high as 40K) is induced. Our findings indicate that the correlation effect plays a crucial role in the iron-based superconductors. (Tl,K)$Fe_xSe_2$, therefore, represents the first Fe-based high temperature superconductor near an insulating AFM.**

All the iron-based superconductors share a common layered structure based on a square planar $Fe^{2+}$ layer [see Fig. 1h], tetrahedrally coordinated pnictogen[19-21] (P, As) or chalcogen[22] (S, Se, Te) anions. It is now widely thought that the interaction that leads to the HTSC originates within these common $Fe^{2+}$ layers, similar in nature to the common $CuO_2$ layer in the cuprate HTSCs. The crystal structure[22] of stoichiometric $TlFe_2Se_2$, which is shown in Fig. 1a, is the same as that of the $BaFe_2As_2$ (122-type iron-pnictide) with the $ThCr_2Si_2$-type. But in the $TlFe_2Se_2$ coumpound, one of the known two 122-type iron-chalcogenide, another $TlFe_2S_2$[23], Tl occuring as a monovalent $Tl^{+1}$ here, there are always Fe vacancies on the Fe square planar[24,25]. So the actual component should be rewritten as $TlFe_xSe_2$ ($1.3\leq x<2.0$). The most interesting is that Fe square layers in $TlFe_xSe_2$ and $TlFe_xS_2$ compounds also exhibit a long-range AFM order. Based on the Mössbauer and neutron diffraction studies, Haggstrom et al.[26] and Sabrowsky et al.[27] argued that the AFM order in $TlFe_xSe_2$ and $TlFe_xS_2$ compounds is associated with the long range order of Fe-vacancy, as shown in Fig. 1 (f) and (g) for $x$=1.5 and 1.6, respectively. In this opinion, the AFM order might be suppressed by increasing the Fe content, in addition, the Fe layer should become a square pattern ($x$=2.0) in the end (Fig. 1h), which exists in all the Fe-based superconductors. The suspecting for superconductivity emergence in this new 122-type iron-chalcogenide motivates us to grow $TlFe_xSe_2$ single crystals with a relatively high Fe-content.

At first, we tried to grow $TlFe_xSe_2$ single crystals with a relatively large $x$ value using various nominal compositions as the starting materials, *i.e.* $Tl_zFe_2Se_2$ ($0.4\leq z\leq 1.0$), $TlFe_zSe_2$ ($1.5\leq z\leq 3$). We have gotten a series of $TlFe_xSe_2$ ($1.3\leq x\leq 1.70$) single crystals. We observed an AFM transition at the Neel temperature $T_N$=365K and 265K, respectively, in the $TlFe_{1.30}Se_2$ and $TlFe_{1.47}Se_2$ crystals. Particularly, we found that the temperature dependence of magnetic susceptibility, $\chi$(T), shows a similar behavior to that in all the parent of the Fe-based compounds[28-33] (see Fig. S1a and S1b). And we also observed that their temperature dependence of *ab*-plane resistivity, $\rho$(T), exhibits a thermally activated behavior with a activation energy $E_a$ of 80.2meV and 57.7meV (see the inset in Fig. S1a and S1b), respectively. Furthermore, superconductivity with a transition temperature 22.4K ($T_c^{mid}$, the middle temperature of the transition) and zero resistivity at 20K ($T_c^{zero}$, the temperature zero resistance) is indeed first observed in the $TlFe_{1.70}Se_2$ crystals (See Fig. S2), but magnetic susceptibility measurement shows its superconducting volume fraction (SVF) being very small (<3%). At the same time, we found that it is difficult to grow a $TlFe_xSe_2$ single crystal with x>1.7.

This indicates that, in order to get bulk superconductivity in this system, it is necessary to increase further the Fe-content. We tried to grow $(Tl,K)Fe_xSe_2$ single crystals by adding K in the starting materials and choosing three group nominal compositions as the starting materials, *i.e.* $Tl_{0.5}K_zFe_2Se_2$(0.15≤z≤0.45), $Tl_{0.4}K_zFe_2Se_2$(0.2≤z≤0.5) and $Tl_{0.8-z}K_zFe_2Se_2$ (0.1≤z≤0.4). Based on the composition analyses (see Table S1) for the grown crystals by an Energy Dispersive X-ray Spectrometer (EDXS), we indeed found that the addition of K element can lead to the increase of the Fe content in the grown crystals, the maximum *x* value can reach to 1.88 (see Table S1). Figure S3 in the Supplementary shows the Fe and K content in the grown crystals as a function of the K content in the starting materials with the nominal composition $Tl_{0.5}K_zFe_2Se_2$(0.15≤z≤0.45). And we also found that the composition of Tl+K in the grown crystals is always of 1.0, regardless of their composition in the starting materials, although the relative ratio between Tl and K depends on the starting composition. A series of $Tl_{1-y}K_yFe_xSe_2$ (1.50≤x≤1.88, 0.14≤y≤0.57) single crystals has been gotten. Figure 1b presents the photos of $Tl_{0.48}K_{0.52}Fe_{1.50}Se_2$ crystals before cleaved and one piece crystal of $Tl_{0.64}K_{0.36}Fe_{1.83}Se_2$.

Figure 1c, 1d and 1e show the single crystal and powder x-ray diffraction (XRD) patterns for the $Tl_{0.64}K_{0.36}Fe_{1.83}Se_2$ and $Tl_{0.48}K_{0.52}Fe_{1.50}Se_2$. All the peaks in the powder XRD pattern can be well indexed with a $ThCr_2Si_2$-type structure (space

group: *I4/mmm*). The lattice parameters, $a$=3.88Å, $c$=14.05Å for $Tl_{0.64}K_{0.36}Fe_{1.83}Se_2$; $a$=3.90Å, $c$=13.99Å for $Tl_{0.48}K_{0.52}Fe_{1.50}Se_2$ crystals were obtained by the fitting XRD data. Broader peaks in the powder XRD patterns might indicate the existence of some disorder in the crystals, and it is difficult to observe super-lattice peaks of Fe vacancy order. Only (*00l*) peaks were observed in the single crystal XRD patterns, indicating that the crystallographic *c* axis is perpendicular to the plane of the single crystal. The single crystal and powder XRD results for other composition samples are similar to these, their lattice parameters are listed in Table S1 in the Supplementary.

Figure 2a and 2b shows the $\rho$(T) for the (Tl,K)$Fe_xSe_2$ (1.50≤x≤1.85) crystals. For the x=1.50 and 1.65 crystals, their resistivity increases rapidly with decreasing temperature and the room temperature resistivity are about 27.8 mΩ·cm and 27.2 mΩ·cm, respectively. Similar to the case of the $TlFe_{1.30}Se_2$ and $TlFe_{1.47}Se_2$ crystals mentioned above, their $\rho$(T) also exhibits the thermally activated behavior: $\rho=\rho_0\exp(E_a/k_BT)$, where $\rho_0$ and $k_B$ are a prefactor and the Boltzmann constant, respectively. The activation energy $E_a$ was estimated to be about 36 meV and 24 meV from the fitting results (see red line for x=1.50 crystal in Fig 2c, in Fig. S1c for x=1.64 crystal), respectively.

The most surprising is that the $\chi$(T) behavior for the x=1.50, and 1.65 insulators, as well as the $TlFe_{1.30}Se_2$ and $TlFe_{1.47}Se_2$ mentioned above (see Fig S1a and S1b), is very similar to that[29] in the $BaFe_2As_2$, which is a parent of 122-type iron-pnictide and a metal. As shown in the inset in Fig. 2c, the susceptibility of the x=1.50 crystal decreases monotonically with temperature and shows a linear temperature dependence above 250K. At 250K the susceptibility shows a rapid decrease which may be ascribed to the occurrence of AFM transition associated with the Fe-vacancy order shown in Fig. 1(f). In addition, the susceptibility at 400K is $3.48\times10^{-3}$ emu/mol, which is the same order of magnitude as that of $BaFe_2As_2$[29]. For the x=1.64 crystal, the susceptibility shows a rapid decrease at 123K, which may be associated with the Fe-vacancy order shown in Fig. 1(g), and high temperature susceptibility exhibits a similar behavior to that of the x=1.50 crystal. These results indicate that the $T_N$ decreases with the increase of the Fe content in the crystal. The observed increase of $\chi$ with $T$ in the FeAs-based compounds has been attributed to itinerant electron antiferromagnetic spin fluctuation[32], or the alternatively to AF correlations between local moments in a strong coupling description[34], as discussed for the cuprate parent compounds[35]. Here, our findings support obviously the latter, because the (Tl,K)$Fe_xSe_2$ (x=1.30, 1.47, 1.50 and 1.64) crystals with a super-lattice of Fe-vacancy are an insulator. Recently, Zhu et al[36]. pointed out by means of theory calculations that the expansion of the Fe square lattice can gives rise to a narrowing of the 3d-bands and a concomitant enhancement of U/t (Here t and U are respectively the characteristic band-width and Coulomb repulsion of the Fe 3d electrons.) pushes the system to the Mott region.

With the *x* value further increasing, the $\rho$ value at the lower temperatures rapidly decreases, for example, at 40K, the *x*=1.64 crystal $\rho$ =18.6 Ω·cm, while the x=1.69 crystal $\rho$ drops to 0.46 Ω·cm. The interesting is that a superconducting transition starts to occur at 25.8K and 27.6K, respectively, for x=1.69 and 1.76 crystal, but no zero resistivity is observed above 2K (see Fig. 2a), indicating that superconductivity starts to emerge and coexists with an insulator phase in the 1.69≤x≤1.76 crystals. In the x=1.78 crystal, a sharp superconducting transition occurs at $T_c^{mid}$ =28.4K, and reach to zero resistivity at $T_c^{zero}$=27.4K, confirmed by the existence of diamagnetism below $T_c$ (See Fig. 2d). For the *x*=1.83, *i.e.*$Tl_{0.64}K_{0.36}Fe_{1.83}Se_2$ crystal, $T_c^{mid}$ increases to 31.1K and $T_c^{zero}$=30K. And for the x=1.84 ($Tl_{0.69}K_{0.31}Fe_{1.84}Se_2$) and 1.85 ($Tl_{0.75}K_{0.25}Fe_{1.85}Se_2$) crystals, although their $T_c^{mid}$ values are almost the same as that in x=1.83, they have a much larger SVF (see Fig.2d). We found that the superconducting transition temperatures, $T_c^{mid}$ and $T_c^{zero}$, vary only with the Fe content in the crystals, and are independent of the K content in the crystals.

In addition, for both the x=1.78 and 1.83 crystals, their $\rho$ value has a maximum at about 75K, where a metal-insulator transition occurs. While for the x=1.85 crystal, although it exhibits a bulk superconductivity below 30K with larger SVF, its resistivity at the normal state is almost independent of *T* and about of 10 mΩ cm. Hall coefficient measurements have confirmed that they are an electron type superconductor (see Fig.4), and the carrier concentration in x=1.83 and 1.85 crystals is almost the same, although there is some differences in $\rho$(T) at the normal state between them. This implies that (Tl,K)$Fe_xSe_2$ (1.78≤x≤1.85) might reside in the under-doping region in the phase diagram. There is a tendency that $T_c$ should increases to the highest one in the crystal with an optimal Fe-content.

We check again the superconducting transition in $\rho$(T) for the (Tl,K)$Fe_xSe_2$ crystals with different x value. As shown in Fig. 3a, for the x=1.78 crystal, there is only a sharp transition at $T_c$=28.4K. While in the x=1.84 crystal, two transitions occur at $T_c$=29.8K and 33.0K, respectively. Most interesting is that three transitions occur at $T_c$=30.4, 34.1, and 40.4K, respectively, in the $Tl_{0.75}K_{0.25}Fe_{1.88}Se_2$ crystal, which is one with the highest Fe content gotten by us. The existence of superconducting phase with $T_c$=40K in the crystals with higher Fe content is also confirmed by $\rho$(T) at various fields for the $Tl_{0.64}K_{0.36}Fe_{1.83}Se_2$ crystal, as shown in Fig. 3d. Compared with the x=1.78 crystal with only one transition at 28.4K, the transition at each magnetic field starts at 40K for the x=1.83 crystal. The multi-transition in $\rho$(T) was observed in all the crystals with x≥1.83, regardless of the K content. This means that the domains with various Fe content, corresponding to different carrier concentration, perhaps exist in the crystals. In the majority domain, superconductivity with $T_c$=30K emerges. The volume fraction of the domains with higher $T_c$ value and higher Fe content, which should be unstable due to the $Fe^{2+}$ valence limitation, may be very small because no diamagnetic signal appears at higher temperature. To our knowledge, this phase separation behavior is first observed in Fe-based superconductors. Superconductivity with $T_c$=40K emergence in the (Tl,K)$Fe_xSe_2$ crystals with relatively high Fe content is reminiscent of the highest $T_c$=38K in the optimal doping $Ba_{1-x}K_xFe_2As_2$ compound[8].

Bulk superconductivity with $T_c$=40K may emerges in the crystal with an optimal Fe-content, but it is difficult to realize in this system due to the $Fe^{2+}$ valence limitation.

In order to explore the effect of the Fe content on the electronic structure in (Tl,K)$Fe_xSe_2$ system, we choose four crystals with $x$=1.84, 1.83, 1.64 and 1.50 to measure Hall coefficient. Their temperature dependence of Hall coefficient, $R_H$(T), is shown in Fig. 4. The negative value of $R_H$ in all the temperature range indicates that the carrier in (Tl,K)$Fe_{2-x}Se_2$ compound is dominated by electrons, which is consistent with the electronic band calculation result for the stoichiometric $TlFe_2Se_2$ reported by Zhang et al.[37] For both the $Tl_{0.48}K_{0.52}Fe_{1.50}Se_2$ and $Tl_{0.53}K_{0.47}Fe_{1.64}Se_2$ crystals, which are an AFM insulator, the $|R_H|$ value increases sharply at $T_N$=250 and 123K, respectively, where an electron localization occurs. This is a typical behavior of the AFM Mott insulator. The increase of the Fe-content results in the enhancement of the carrier concentration ($|R_H|$ value decrease), which can be regards as a self-doping of $Fe^{2+}$ electron to the Fe layer with a super-lattice of Fe vacancy.

We have constructed a phase diagram of the magnetism and superconductivity for (Tl,K)$Fe_xSe_2$ with $1.3 \le x \le 2.0$, which is shown in Fig. 5. There are three composition regions with distinct physical properties. For the region of $1.3 \le x < 1.7$, the compound is an AFM insulator, which may be associated with the existence of various super-lattice of Fe-vacancy. The Néel temperature, $T_N$, decreases with increasing Fe content. In the region of $1.70 \le x < 1.78$, a superconducting transition was observed in $\rho$(T), but superconducting volume fraction is less than 1% estimated from the susceptibility data. An inhomogeneity of Fe-vacancy distribution may result in the coexistence of superconductivity and insulator phase. In the region of $1.78 \le x \le 1.88$, bulk superconductivity emerges. The observation of multi-transition in $\rho$(T) for the crystals with $x \ge 1.83$ implies that phase separation may occur in the crystals and an optimal Fe content for bulk superconductivity with the highest $T_c$=40K does not reach due to the limitation of Fe valence. Correspondingly, the phase diagram of (Tl,K)$Fe_xSe_2$ system appears very similar to that of the cuprate superconductors.

In summary, we first report the measurements of resistivity, magnetic susceptibility and Hall coefficient for (Tl,K)$Fe_xSe_2$ crystals. We found that (Tl,K)$Fe_xSe_2$ ($1.3 \le x < 1.7$) crystals are antiferromagnetic insulators. With the increase of the Fe content, bulk superconductivity with $T_c^{mid}$=31K occurs in (Tl,K)$Fe_xSe_2$ ($1.78 \le x \le 1.88$) crystals, and the onset superconducting transition temperature $T_c^{onset}$ goes up as high as 40K. It is found that the phase diagram of (Tl,K)$Fe_xSe_2$ system is very similar to that of the cuprate superconductors. The evolution from an AFM insulating state to a superconducting state can be driven by increasing Fe content in (Tl,K)$Fe_xSe_2$, making these compounds a promising new platform for further exploration of the electron correlation effect. At the same time, our results show that orderly introducing Fe vacancy represents a means to induce Fe-based parents to the Mott insulating part of the phase diagram.

**Method Summary**

(Tl,K)$Fe_xSe_2$ single crystals were grown using a self-flux method. The starting materials $Tl_2Se$, $K_2Se$, Fe and Se powder were mixed according to a various nominal composition in a glove box and loaded in a thin corundum crucible. Then the crucible was sealed in an evacuated quartz tube. The sealed silica tube was slowly heated to 950ºC and held for 6 hours. Then let temperature decrease to 700ºC in a rate of 3ºC/hrs, and finally cool down to room temperature by switching off the furnace. In each step to prepare sample, it is very important to manage carefully due to $Tl_2Se$ poison. The composition of crystals was determined using an Energy Dispersive X-ray Spectrometer (EDXS). The electrical resistivity and Hall coefficient were measured on a Physical Properties Measurement System (PPMS; Quantum Design). The dc magnetic susceptibility was measured by using SQUID (Quantum Design).

**Acknowledgements:** This work is supported by the Nature Science Foundation of China, the National Basic Research Program of China (973 Program), and PCSIRT of the Ministry of Education of China. We acknowledge Qimiao Si and Jianhui Dai for discussion and encouragement.

**Author Contributions:** M.H.F. designed this study, analysed the data, did partially the experiments and wrote the paper. H.D.W. did the main experiments and analysed the data. C.H.D., Z.J.L. and C.M.F measured the resistivity and prepared partially samples. J.C. measured partially susceptibility. H.Q.Y provided experimental support.

**Author Information:** The authors declare that they have no competing financial interests. Correspondence and requests for materials should be addressed to M.H.F (mhfang@zju.edu.cn)

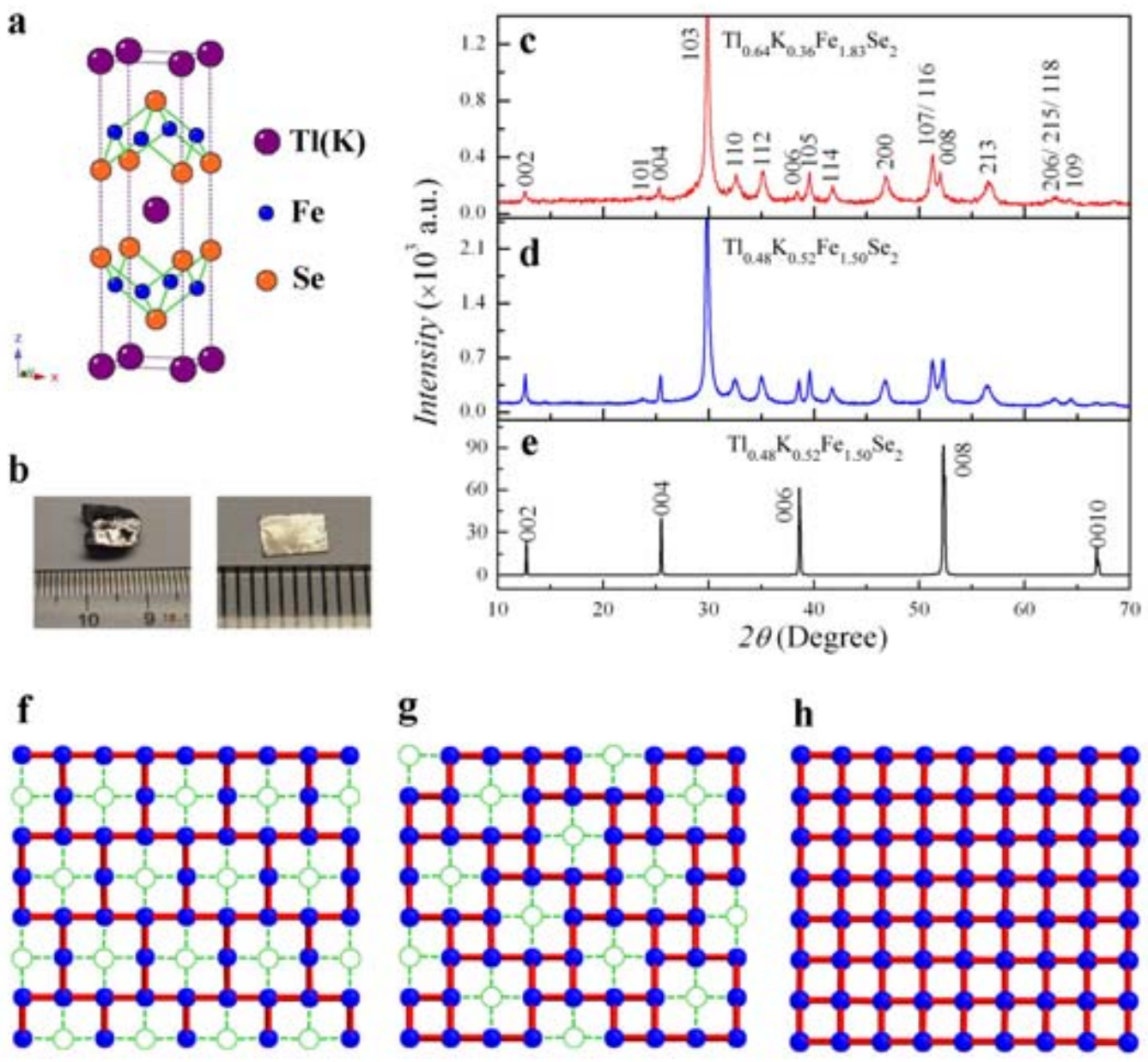

Figure 1 | **Crystal structure and super-lattice of the Fe-vacancy in (Tl,K)$Fe_xSe_2$.** **a,** Crystal structure of stoichiometric $Tl_{1-y}K_yFe_2Se_2$. Tl/K and Se are fully occupied sites of the tetragonal $ThCr_2Si_2$ structure. Fe are partially occupied sites, there are always vacancies partial occupied Fe site due to the requirement of charge-balance as $(Tl,K)^{1+}Fe^{2+}{}_xSe^{2-}{}_2$. We found that for $TlFe_xSe_2$ system, x≤1.7, for $Tl_{1-y}K_yFe_xSe_2$ system (y≠0), x≤1.88, which provides a chance to introduce electrons by modifying the Fe content. **b,** Photos of $Tl_{0.48}K_{0.52}Fe_{1.50}Se_2$ crystals before cleaved and one piece crystal of $Tl_{0.64}K_{0.36}Fe_{1.83}Se_2$. **c,** Powder x-ray diffraction (XRD) pattern of $Tl_{0.64}K_{0.36}Fe_{1.83}Se_2$, the powder was obtained by grinding many pieces of crystals. **d,** Powder XRD pattern of $Tl_{0.48}K_{0.52}Fe_{1.50}Se_2$. **e,** Single crystals XRD pattern of $Tl_{0.48}K_{0.52}Fe_{1.50}Se_2$. **f,** Possible Fe-vacancy order of (Tl,K)$Fe_{1.5}Se_2$. Blue solid circle denotes: Fe atom, Green hollow circle denotes: Vacancy. The iron atoms have two or three iron neighbors. **g**, Possible Fe-vacancy order of (Tl,K)$Fe_{1.6}Se_2$, all iron atoms have three identical neighbors. Both patterns were suggested by M. Zabel et al.[24, 25], Sabrowsky et al.[26] and Haggstrom et al.[27] for the explanation of AFM in $TlFe_xSe_2$ and $TlFe_xS_2$ compounds, based on their Mössbauer and neutron diffraction studies. **h,** $Fe^{2+}$ square planar for stoichiometric (Tl,K)$Fe_2Se_2$, which exists in all the Fe-based superconductors. All iron atoms have four iron neighbors. In (Tl,K)$Fe_xSe_2$, the number of iron nearest neighbors varies with composition.

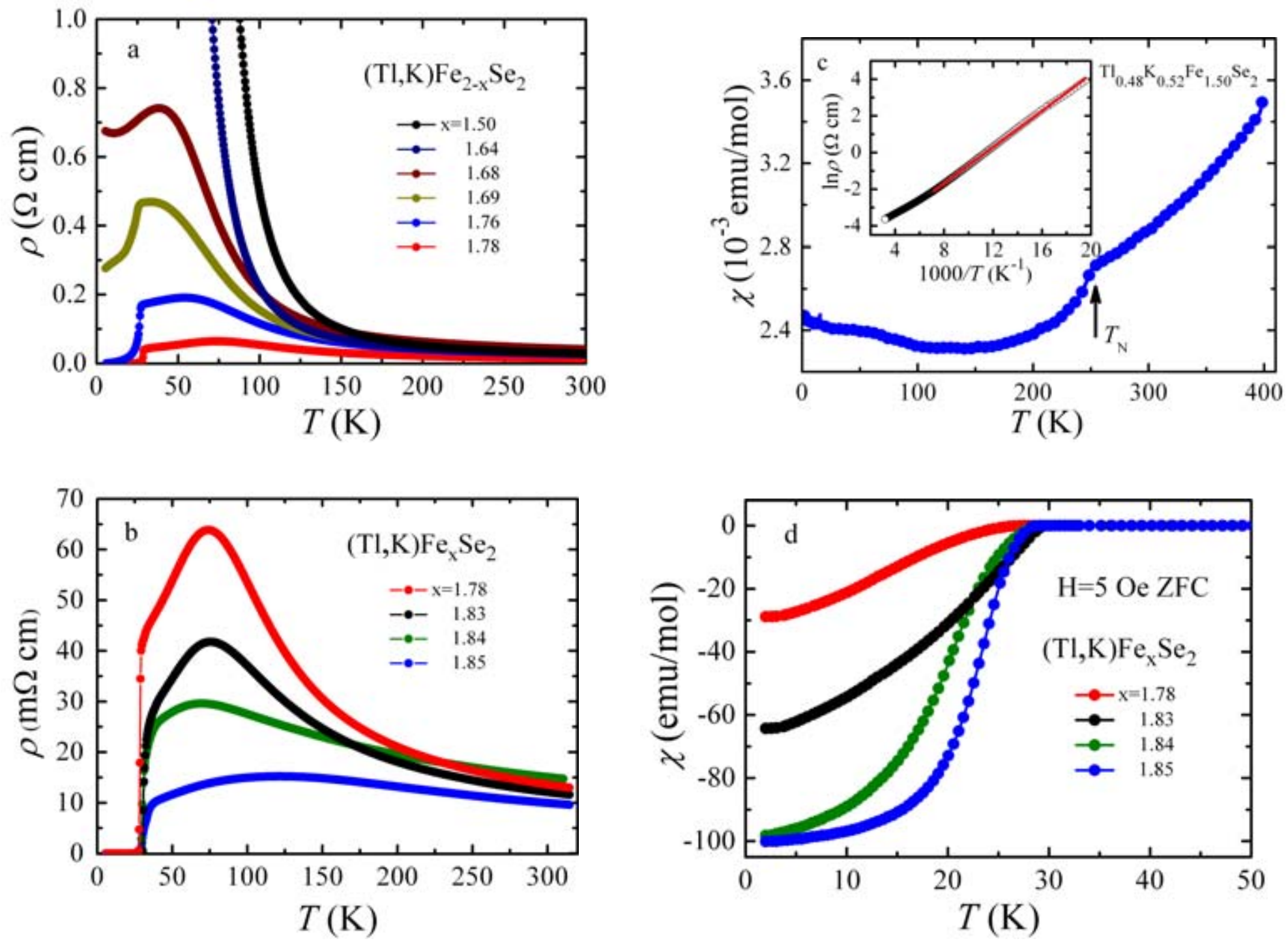

Figure 2 | **Resistivity and Magnetic susceptibility of (Tl,K)$Fe_xSe_2$ (1.50≤x≤1.85). a,** In-plane resistivity $\rho$(T) as a function of temperature for (Tl,K)$Fe_xSe_2$ (x=1.50, 1.64, 1.68, 1.69, 1.76 and 1.78) crystals. **b**, $\rho$(T) for (Tl,K)$Fe_xSe_2$ (x=1.78, 1.83, 1.84, and 1.85) crystals. **c,** The temperature dependence of magnetic susceptibility, $\chi$(T), measured at 1Tesla magnetic field with field cooling process. The inset is ln$\rho$ as a function of 1/*T* for $Tl_{0.48}K_{0.52}Fe_{1.50}Se_2$ crystal. **d,** Magnetic susceptibility data measured with a zero-field cooling history and a field of 5Oe for (Tl,K)$Fe_xSe_2$ (x=1.78, 1.83, 1.84 and 1.85) crystals. Both magnetic susceptibility and resistivity show that bulk superconductivity occurs in the crystals with x≥1.78, and the superconducting volume fraction increases with the increase of x value. The $\chi$(T) behavior for the x=1.50 crystal, which is a thermally-activated semiconductor with an activation energy $E_a$ of 36 meV, is very similar to that in the $BaFe_2As_2$, which is a parent of 122-type iron-pnictide and a metal.

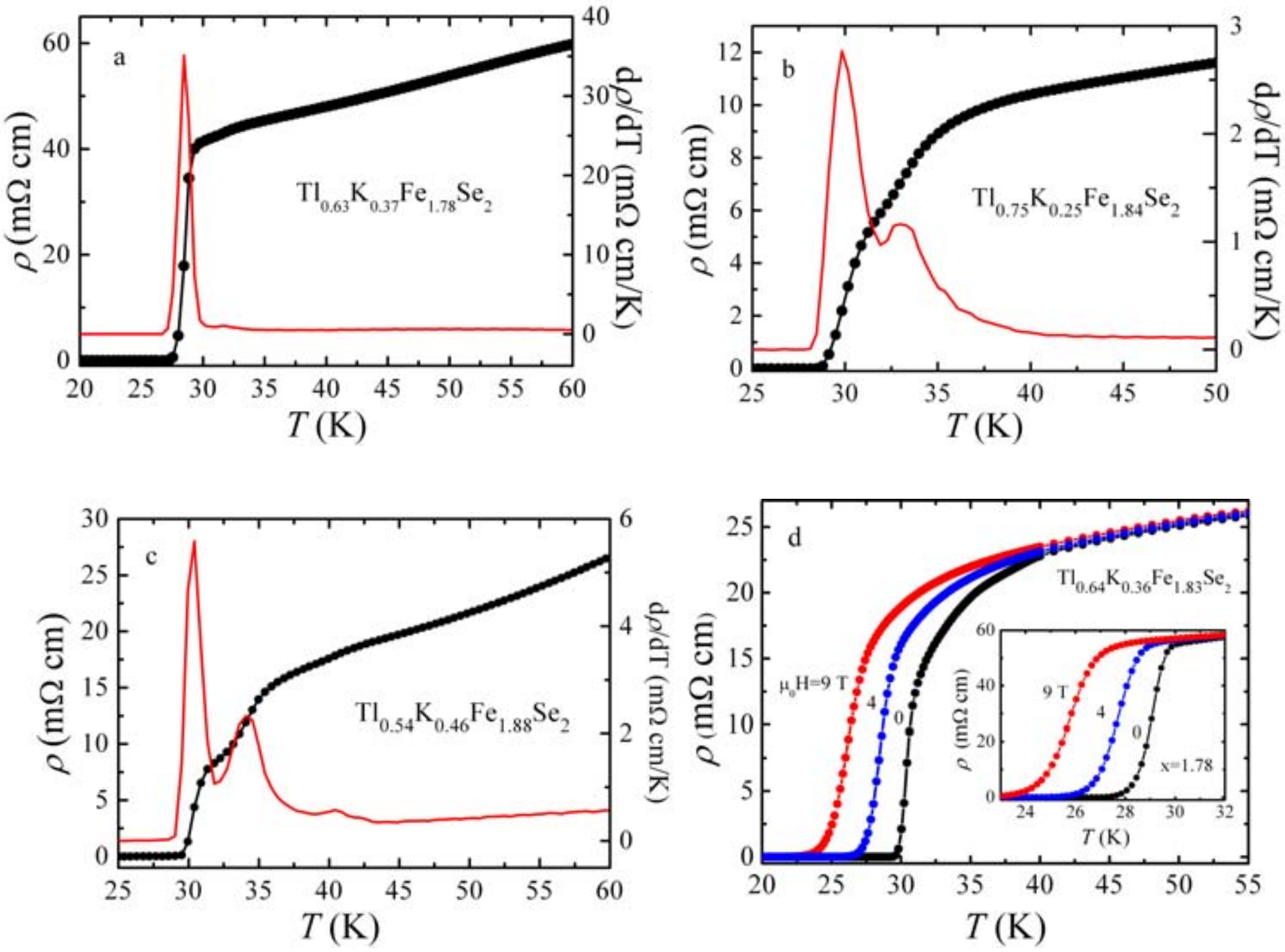

Figure 3 | **The superconducting transition of four typical (Tl,K)$Fe_{1.78}Se_2$ crystals. a,** The temperature dependence of *ab*-plane resistivity, $\rho$(T), and its differential d$\rho$(T)/dT near the superconducting transition for $Tl_{0.63}K_{0.37}Fe_{1.78}Se_2$ crystal. Only one transition appears at $T_c^{mid}$=28.4K. **b,** $\rho$(T) and its differential d$\rho$(T)/dT for $Tl_{0.75}K_{0.25}Fe_{1.84}Se_2$ crystal. Two transitions occur at $T_c$=29.8K and 33.0K, respectively. **c**, $\rho$(T) and its differential d$\rho$(T)/dT for $Tl_{0.54}K_{0.46}Fe_{1.88}Se_2$ crystal, which is one with the highest Fe content gotten by us. Three transitions occur at $T_c$=30.4, 34.1, and 40.4K, respectively. **d,** $\rho$(T) near the superconducting transition in magnetic field of 0, 4, and 9 Tesla for $Tl_{0.64}K_{0.36}Fe_{1.83}Se_2$ crystal, the transition starts at 40K. The inset shows the $\rho$(T) at 0, 4, and 9 Tesla for $Tl_{0.63}K_{0.37}Fe_{1.78}Se_2$ crystal, which has only one transition. The multi-transition in $\rho$(T) was observed in all the crystals with x≥1.83, regardless of the K content.

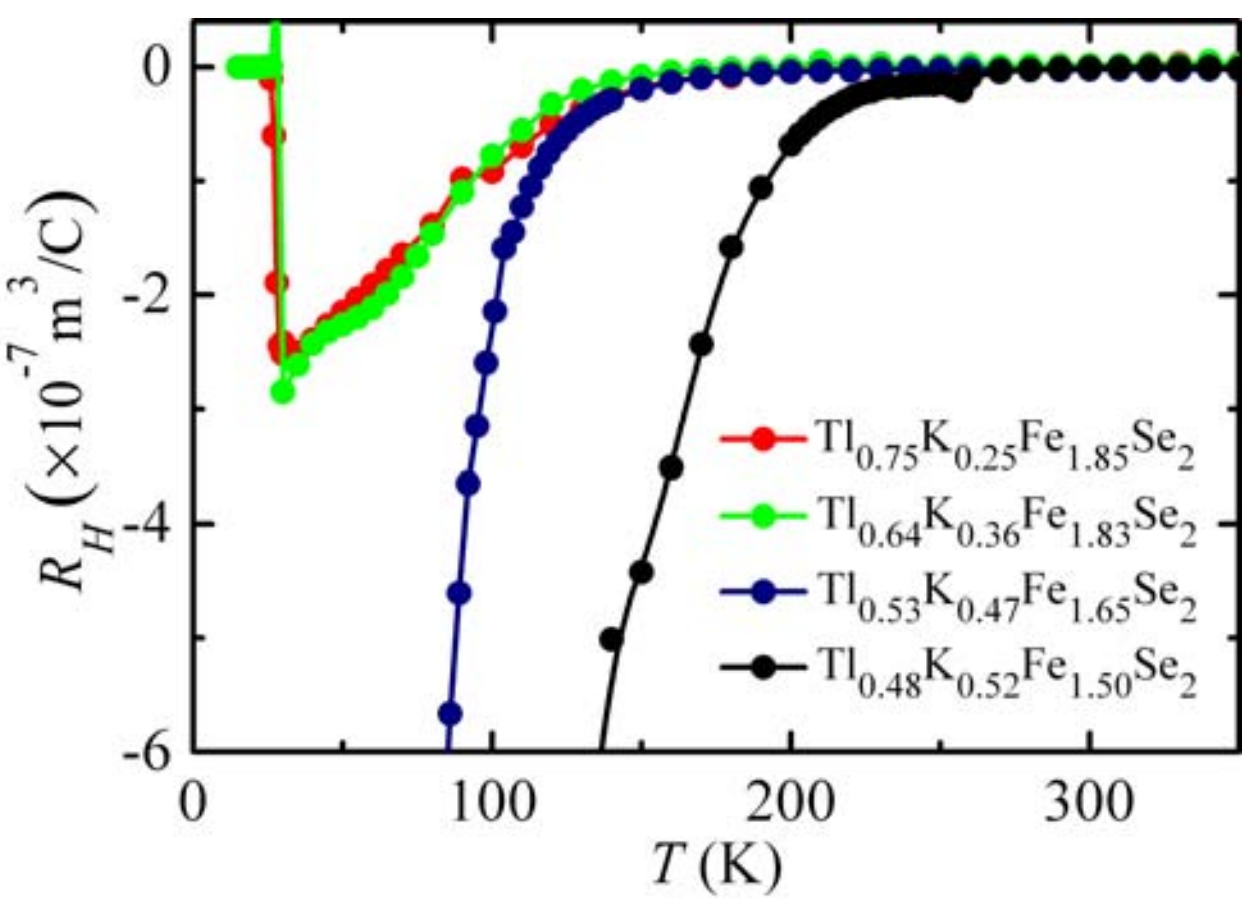

Figure 4 | **The temperature dependence of Hall coefficient, $R_H$(T), for (Tl,K)$Fe_xSe_2$ (x=1.50, 1.64, 1.83 and 1.84) crystals.** The negative value of $R_H$ indicates the carriers dominated by electrons in (Tl,K)$Fe_{2-x}Se_2$ compound. For both $Tl_{0.48}K_{0.52}Fe_{1.50}Se_2$ and $Tl_{0.53}K_{0.47}Fe_{1.64}Se_2$ crystals, the $|R_H|$ value increases sharply at $T_N$=250 and 123K, respectively, indicating that an electron localization occurs at AFM transition temperature. The increase of the Fe-content results in the enhancement of the carrier concentration ($|R_H|$ value decrease), which can be regards as a self-doping of $Fe^{2+}$ electron to the Fe layer with a super-lattice of Fe vacancy. The $R_H$ value seems to be independent of the K content in the crystals.

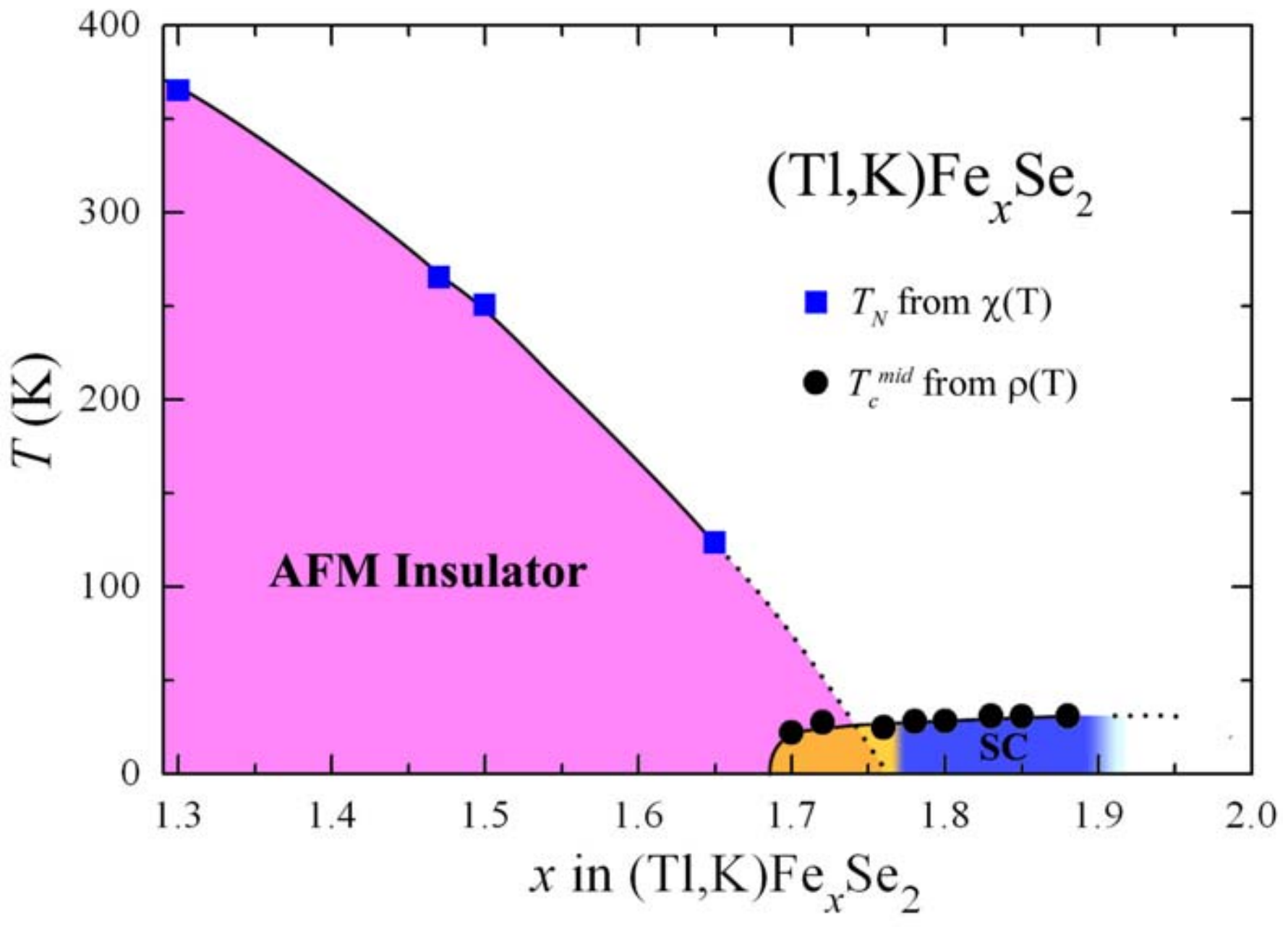

Figure 5 | **The phase diagram of the magnetism and superconductivity for (Tl,K)$Fe_xSe_2$ (1.30≤x≤1.88).** The Néel temperature, $T_N$, of the AFM phase, is determined by the onset temperature of the transition in susceptibility $\chi$(T). The superconducting transition temperature, $T_c^{mid}$, is determined by the middle temperature of the transition in $\rho$(T). In the 1.3≤x<1.7 region, the compound is an AFM insulator, which may be associated with the existence of various super-lattice of Fe-vacancy. In the region of 1.70≤x<1.78, a superconducting transition was observed in $\rho$(T), but superconducting volume fraction is less than 3% estimated from the susceptibility data. An inhomogeneity of Fe-vacancy distribution may result in the coexistence of superconductivity and insulator phase. In the region of 1.78≤x≤1.88, bulk superconductivity emerges. The observation of multi-transition in $\rho$(T) for the crystals with x≥1.83 implies that phase separation may occurs in the crystals. And an optimal Fe content for bulk superconductivity with the highest $T_c$=40K is difficult to get in this system due to the limitation of Fe valence.

**Supplementary Information:**

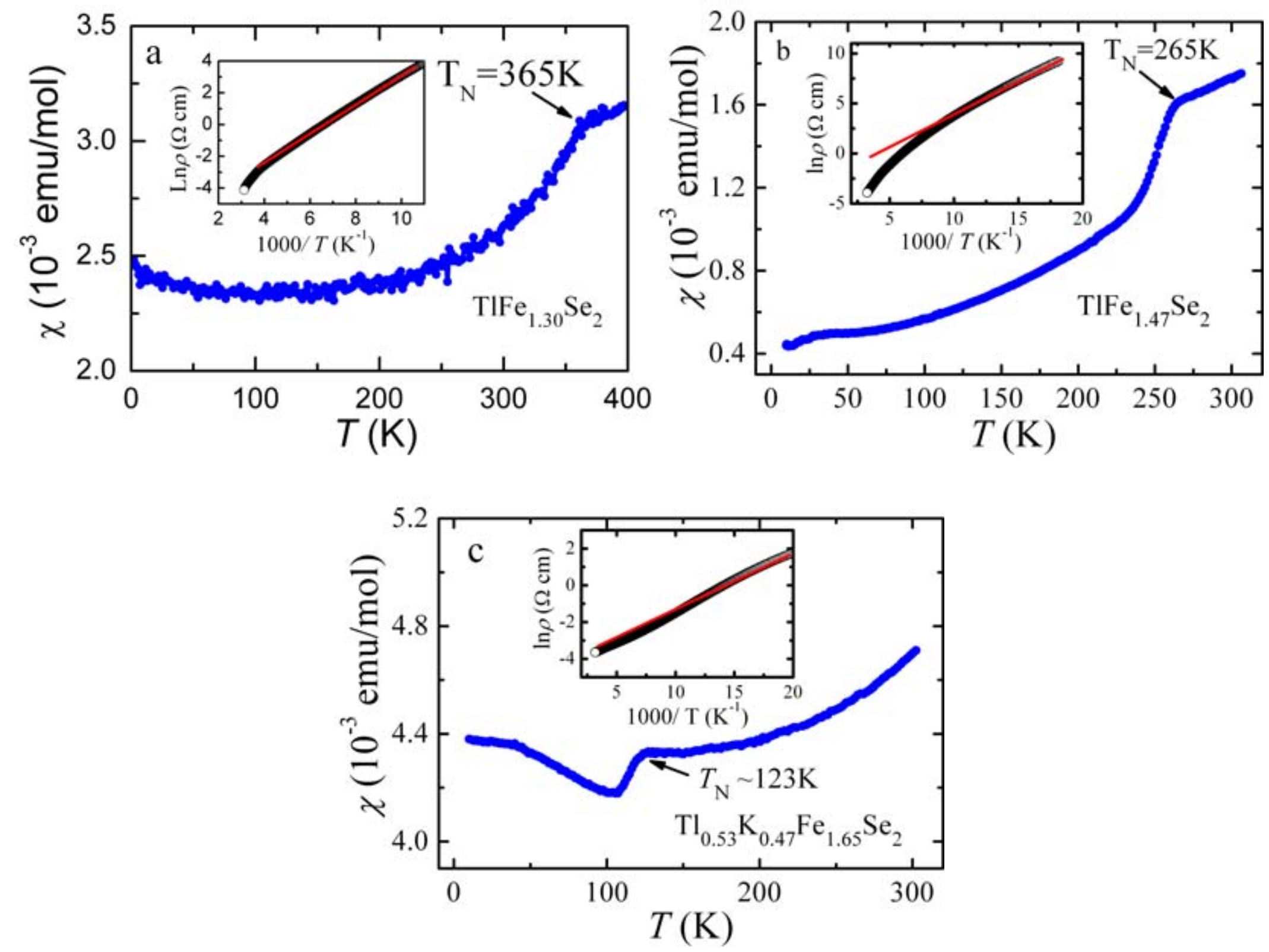

Figure S1| **Temperature dependence of magnetic susceptibility, $\chi$(T), and ln$\rho$ as a function of 1/*T* for three insulators. a,** For $TlFe_{1.30}Se_2$ crystal. $\chi$(T), measured at 1Tesla magnetic field. The inset is the ln$\rho$ versus 1/*T*, an activation energy $E_a$ of 80.2meV was estimated from the fitting as the red line. **b,** For $TlFe_{1.47}Se_2$ crystal. $\chi$(T), measured at 1Tesla magnetic field. The inset is the ln$\rho$ versus 1/*T*, an activation energy $E_a$ of 57.7meV was estimated from the fitting as the red line. **c,** For $Tl_{0.53}K_{0.47}Fe_{1.47}Se_2$ crystal. $\chi$(T), measured at 1Tesla magnetic field. The inset is the ln$\rho$ versus 1/*T*, an activation energy $E_a$ of 24meV was estimated from the fitting as the red line. At low temperature, $\chi$ value shows an upturn, presumably due to small amount impurities in the crystal.

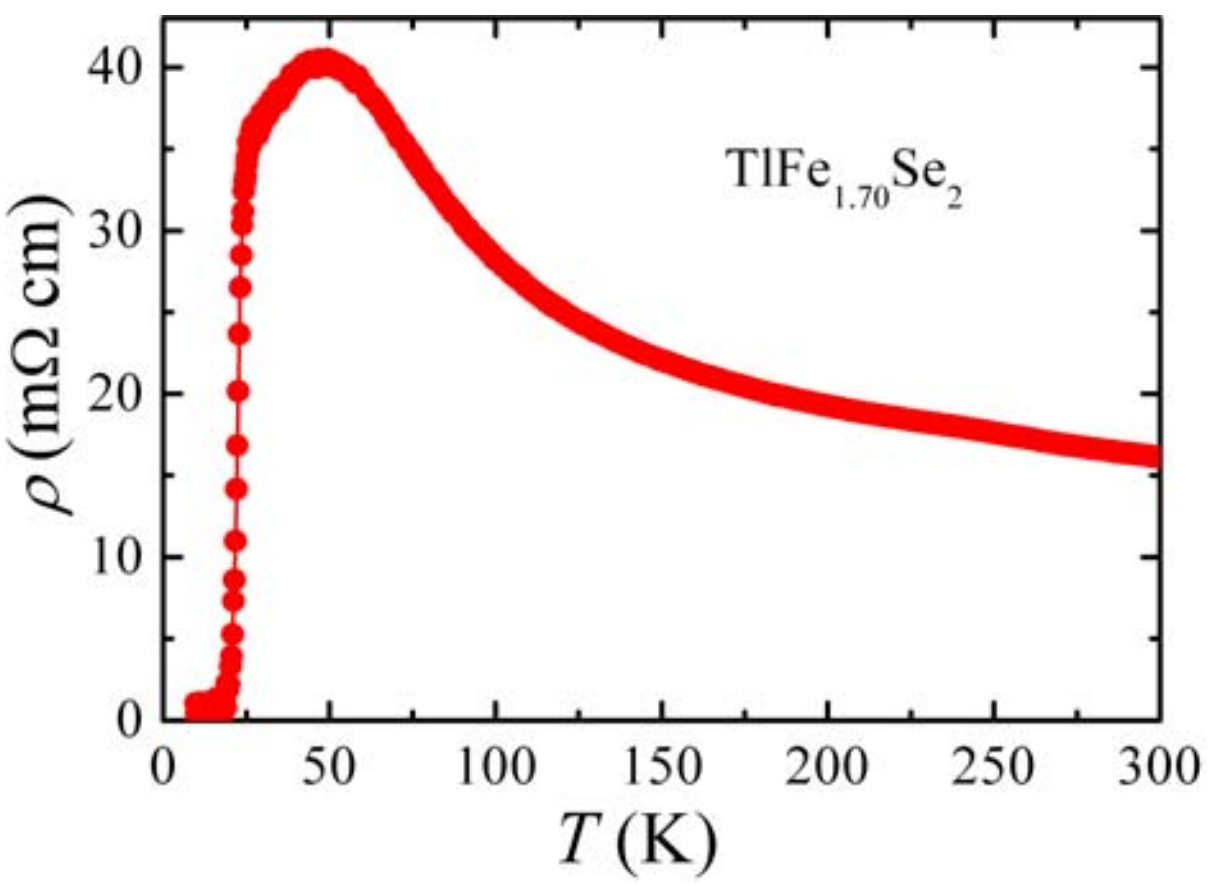

Figure S2 | **The temperature dependence of *ab*-plane resistivity, *ρ*(T), for $TlFe_{1.70}Se_2$ crystal.** Above 50K, *ρ*(T) exhibits a semiconducting behavior, *i.e.* *ρ* increases with decreasing temperature. A superconducting transition occurs at $T_c^{onset}$=24.3K, $T_c^{mid}$=22.4K and $T_c^{zaro}$=20K.

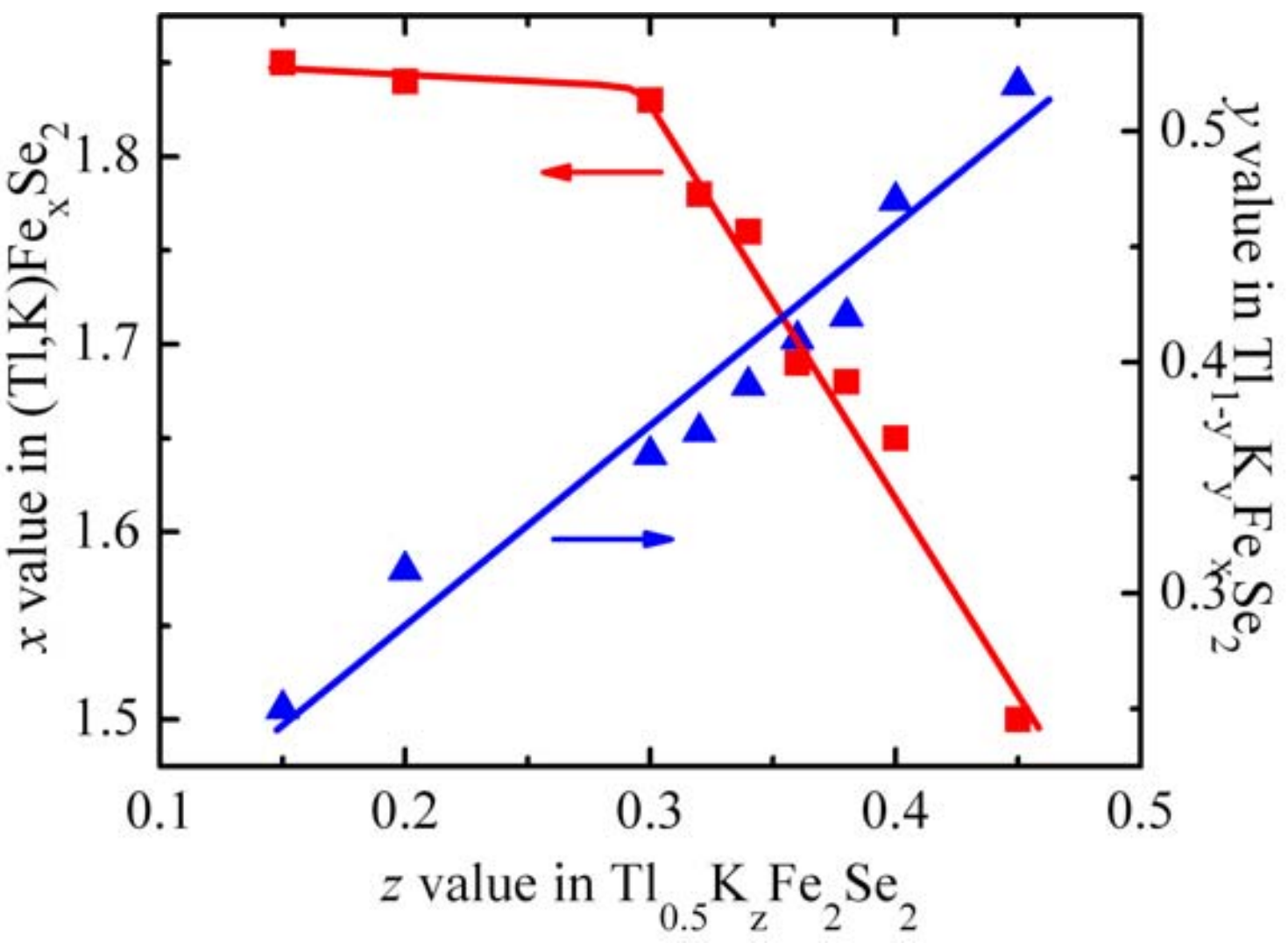

Figure S3 | **The Fe and K contents in the grown crystal as a function of K content, z value, in the starting materials with a nominal composition $Tl_{0.5}K_zFe_2Se_2$.** The K content in the grown crystals increases linearly with increasing K content in the starting materials. When 0.3≤z≤0.45, the Fe content in the grown crystals increases near linearly with the decreasing of K content in the starting material, and goes almost up to a saturation value of 1.83 at z=0.3.

Table S1 | **Nominal composition in the starting materials, average composition for the grown crystals determined by EDXS, lattice parameters *a* and *c* values for crystals from the fitting XRD data, Neel temperature, $T_N$, determined by the onset temperature of transition in $\chi$(T), superconducting transition, $T_c^{mid}$ and $T_c^{zero}$ determined by $\rho$(T) data.**

| Nominal composition | Average composition of the grown crystal | *a* value (Å) | *c* value (Å) | $T_N$ (K) | $T_c^{mid}$ (K) | $T_c^{zero}$ (K) |
|---|---|---|---|---|---|---|
| $TlFe_{1.60}Se_{2.2}$ | $TlFe_{1.30}Se_2$ | 3.90 | 13.89 | 365 | | |
| $TlFe_{1.50}Se_2$ | $TlFe_{1.47}Se_2$ | 3.90 | 13.95 | 265 | | |
| $Tl_{0.80}Fe_2Se_2$ | $TlFe_{1.70}Se_2$ | | | | 22.4 | 20.0 |
| $Tl_{0.4}K_zFe_2Se_2$ | | | | | | |
| z=0.20 | $Tl_{0.69}K_{0.31}Fe_{1.87}Se_2$ | | | | 31.2 | 27.9 |
| 0.30 | $Tl_{0.61}K_{0.39}Fe_{1.86}Se_2$ | | | | 29.2 | 28.2 |
| 0.40 | $Tl_{0.53}K_{0.47}Fe_{1.80}Se_2$ | | | | 28.5 | 27.7 |
| $Tl_{0.8-z}K_zFe_2Se_2$ | | | | | | |
| z=0.10 | $Tl_{0.86}K_{0.14}Fe_{1.72}Se_2$ | | | | 27.9 | <2K |
| 0.20 | $Tl_{0.74}K_{0.26}Fe_{1.83}Se_2$ | | | | 31.2 | 28.9 |
| 0.30 | $Tl_{0.64}K_{0.36}Fe_{1.83}Se_2$ | | | | 31.1 | 30.1 |
| 0.40 | $Tl_{0.53}K_{0.47}Fe_{1.80}Se_2$ | | | | 28.5 | 27.7 |
| $Tl_{0.3}K_{0.3}Fe_2Se_2$ | $Tl_{0.54}K_{0.46}Fe_{1.88}Se_2$ | | | | 31.3 | 29.5 |
| $Tl_{0.5}K_zFe_2Se_2$ | | | | | | |
| z=0.15 | $Tl_{0.75}K_{0.25}Fe_{1.85}Se_2$ | 3.88 | 14.06 | | 31.0 | 28.8 |
| 0.20 | $Tl_{0.69}K_{0.31}Fe_{1.84}Se_2$ | 3.88 | 14.05 | | 30.4 | 28.9 |
| 0.30 | $Tl_{0.64}K_{0.36}Fe_{1.83}Se_2$ | 3.87 | 14.05 | | 29.2 | 28.2 |
| 0.32 | $Tl_{0.63}K_{0.37}Fe_{1.78}Se_2$ | | | | 28.4 | 27.4 |
| 0.34 | $Tl_{0.61}K_{0.39}Fe_{1.76}Se_2$ | 3.80 | 14.06 | | 25.1 | <2K |
| 0.36 | $Tl_{0.59}K_{0.41}Fe_{1.69}Se_2$ | | | | | |
| 0.38 | $Tl_{0.58}K_{0.42}Fe_{1.68}Se_2$ | | | | | |
| 0.40 | $Tl_{0.53}K_{0.47}Fe_{1.65}Se_2$ | 3.89 | 14.06 | 123 | | |
| 0.45 | $Tl_{0.48}K_{0.52}Fe_{1.50}Se_2$ | 3.91 | 13.99 | 250 | | |